\documentclass[%
 reprint,
 amsmath,amssymb,
 aps,
]{revtex4-2}

\usepackage{graphicx}% Include figure files
\usepackage{amsmath}
\usepackage{dcolumn}% Align table columns on decimal point
\usepackage{bm}% bold math
\begin{document}

\preprint{APS/123-QED}

\title{Unimodular-like times, evolution and Brans-Dicke Gravity}

\author{Paolo M Bassani}
\thanks{paolo.bassani22@imperial.ac.uk}

\author{João Magueijo}
\thanks{j.magueijo@imperial.ac.uk}

\affiliation{%
 Theoretical Physics Group, The Blackett Laboratory, Imperial College, Prince Consort Rd., London, SW7 2BZ, United Kingdom\\}%
 
\date{\today}% It is always \today, today,

\begin{abstract}
In unimodular-like theories, the constants of nature are demoted from pre-given parameters to phase space variables. Their canonical duals provide physical time variables. We investigate how this interacts with an alternative approach to varying constants, where they are replaced by dynamical scalar fields. Specifically we investigate the Brans-Dicke theory of gravity and its interaction with clocks dual to the cosmological constant, the Planck mass, etc. We crucially distinguish between the different role of Newton's $G$ in this process, leading to the possibility of local Lorentz invariance violation. A large number of possible theories emerge, for example where the Brans-Dicke coupling, $\omega$, depends on unimodular-like times (in a generalization of scalar-tensor theories), or even become the dual variable to unimodular-like clocks ticking variations in other demoted constants, such as the cosmological constant. We scan the space of possible theories and select those most interesting regarding the joint variations of the Brans-Dicke $\omega$ and other parameters, (such as the cosmological constant); and also regarding their energy conservation violation properties. This ground work is meant to provide the formalism for further developments, namely regarding cosmology, black holes and the cosmological constant problem. 
\end{abstract}

\maketitle

%\tableofcontents

\section{INTRODUCTION}
In this paper we consider the result of cross-breeding two major strands in the field of varying constants, pioneered by Dirac~\cite{dirac}. On the one hand we have the Brans-Dicke-like approach, in which constants are replaced by dynamical scalar fields. This approach originally targeted Newton's constant $G$~\cite{BD}, but was subsequently applied to the electron charge~\cite{Bek,BSM}, the electron mass~\cite{emass}, the speed of light~\cite{CovVSL,eorc}, the electro-weak theory parameters~\cite{dagny}, and many others. On the other hand we have the unimodular-like approach, based on the unimodular theory of gravity~\cite{unimod1,unimod,UnimodLee1,alan,daughton,sorkin1,sorkin2,Bombelli,UnimodLee2}
specifically
in the Henneaux-Teitelboim (HT) formulation~\cite{unimod}. It allows for the constants of nature to be demoted from pre-given fixed parameters to phase space variables which are constant as a result of the equations of motion (i.e. constants-on-shell only, or integration constants). The original target for this exercise was the cosmological constant, $\Lambda$, but the procedure has been generalized to the Planck mass~\cite{pad,pad1,lomb,weinberg,padilla}, the speed of light and of gravity~\cite{evolution}, and the gravitational coupling~\cite{vikman,vikman1,vikman2}. Even though the implications for quantum theory have been understood for a while~\cite{Hawking3f} (and also more recently~\cite{gielen,gielen1,JoaoLetter,JoaoPaper}), it was only in the recent past that it was realized that such theories can become varying constant theories too~\cite{evolution}. This is because
the demoted ``constants'' acquire canonical duals that can be used as time variables~\cite{unimod,sorkin1,sorkin2,Bombelli,UnimodLee2,JoaoLetter,JoaoPaper}. Time-independence of the Hamiltonian with respect to such duals leads to on-shell constancy of the ``constants'', but reciprocally, time-dependence, or evolution, leads to their variation. 

These two strands have very different motivations and technical structures. For example, the second approach is designed so as not to produce modifications in the Einstein equations and no new terms in the stress energy tensor. This is achieved by not introducing the metric in the new terms added to the action, in an ingenious prescription that does not break diffeomorphism invariance. This is in marked contrast with the first approach, which does produce corrections to the field equations, in the form of a contribution to the stress energy tensor (similar to that of any dynamical scalar field), but not only (for example the failed ``boundary'' terms appearing in the gravitational equations of Brans-Dicke theory~\cite{BD}). 

There are good reasons for investigating a combination of the two approaches. For example, the $\omega$ coupling defined in Brans-Dicke theory could become a function of the time variables defined in unimodular-like theories. 
Specifically, in standard unimodular theory the dual to $\Lambda$ is a time variable which on-shell becomes a physical definition of time: 4-volume time; what if $\omega$ depends on this time? Likewise, $\omega$ could be a target for the unimodular-like approach, leading to a dual ``Brans-Dicke'' time. Should there be a dependence on this time, $\omega$ would vary. Both approaches  generalize and change the phenomenology of scalar-tensor theories. Furthermore, such hybrid theories lead to inter-linked variations of $G$ and other constants, such as $\Lambda$. 
%In summary, a combination of the two approaches may amount for more than the sum of the parts. 
%On a more fundamental level, we may wonder 
The implications for scenarios seeking to stabilize the vacuum energy, such as the sequester~\cite{pad,pad1,lomb,weinberg,padilla} are obviously interesting. 
In plain GR there are not many parameters to play with: only $\Lambda$ and $G$. The addition of Brans-Dicke parameters could be a game changer. 
%For all these reasons a combination of the two paradigms for varying constants amounts to more than the sum of the parts. 

%(differentiating between the Planck mass and the gravitational coupling [CITE] if need be). Thus, the sequestration game is played with them. The addition of $\omega$, in the various interconnections sketched above, could change the game. 

Naturally, in our cross-breeding project we are faced with several forks, leading to a large number of possibilities. There is one point of intersection between the two paradigms: a Brans-Dicke scalar without a kinetic term ($\omega=0$) and a unimodular-like theory targeting the Planck mass only.
%~\footnote{We really have to comment on the EC vs EH fork here or this might be confusing for people in the know.}. 
This case reduces to work already done~\cite{evolution}. Nonetheless the local Lorentz invariance violation (LLIV) possibilities opened up in~\cite{evolution} allow for variations. For example, the $G$ appearing in the gravity canonical pairs (and carrying over to the commutators, upon quantization) could be kept fixed, while the $G$ multiplying the  gravitational Hamiltonian (and other constraints) be allowed to vary~\cite{JoaoLetter,JoaoPaper}. Such a separation of $G$s according to their function follows the philosophy of~\cite{vikman,vikman1}. Obviously, this breaks Lorentz invariance, leading to a version of Brans-Dicke theories we will label LIVBD theories. We can then extend this approach to dynamical theories ($\omega\neq 0$); or not. Thus, we find at least two forks: dynamical vs non-dynamical (in the sense of $\omega\neq 0$ vs. $\omega=0$), and LLIV vs non-LLIV.

This paper is structured as follows. In Section~\ref{review} we start by reviewing the two paradigms this paper is based on. Then in Section~\ref{reduct} we combine them and reduce the action to the framework assumed for the rest of the paper: reduction to spatial homogeneity and isotropy. We then start scanning through the possibilities we found most interesting, given the forks described. In Section~\ref{forks} we explain in more technical detail what these forks entail. We then subject these theories to the further forks already presented in~\cite{evolution}. First we have to decide which ``constants'' to target and which should produce clocks, which should vary as a function of these clocks. These further forks lead to an exponential explosion in the number of possible theories, and in this paper we will select those that we found the most fertile regarding the interlinked variations of $\omega$ and other parameters, and the source terms appearing in the energy conservation equation, fundamental to early Universe modelling. 
%referring to [JM: MSc Thesis, which you should put in the arxives] for the gold sieving work behind it. ....

\section{Review of previous work}\label{review}
We first review the two approaches mentioned in the introduction, starting with the Henneaux-Teitelboim formulation of unimodular gravity, and how it can be generalized to constants other than $\Lambda$. 

\subsection{The unimodular-like approach}

%Postulating the variability of physical constants, it is necessary to provide parameters for their evolution. Therefore, we propose to use physical, relational times dual to every constant we will consider. 
%Following the Henneaux-Teitelboim (HT) formalism of unimodular gravity [REF HT], a time dual to the cosmological constant $\Lambda$ can be obtained, called the 4-volume or unimodular time [REF].
The HT formulation~\cite{unimod} of unimodular gravity~\cite{unimod1,unimod,UnimodLee1,alan,daughton,sorkin1,sorkin2,Bombelli,UnimodLee2} introduces an additional term $S_U$ to the base gravity action $S_0$, which, usually, is the Einstein-Hilbert action. This additional term leads to an overall action of the form:
\begin{equation}
     S_0 \rightarrow S = S_0 + S_U = S_0 - \int{d^4 x \: \rho_\Lambda \: \partial_\mu \mathcal{T^{\mu}}} \label{unimodular_1}
\end{equation}
where $\rho_\Lambda$ is the cosmological constant {\it density} and $\mathcal{T^{\mu}}$ is a vector density, such that $S_U$ is fully diffeomorphism invariant (this happens because, for vector densities, $\nabla_\mu = \partial_\mu$, and because the integral is a scalar). Since the $S_U$ term does not depend on the metric, the Einstein field equations remain unchanged. Furthermore, varying action \eqref{unimodular_1} with respect to $\mathcal{T^{\mu}}$ and $\rho_\Lambda$, we obtain the equation of motion for the cosmological constant and the unimodular constraint:
\begin{equation}
    \frac{\delta S}{\delta \mathcal{T}^\mu_\Lambda} = \frac{\delta S_U}{\delta \mathcal{T}^\mu_\Lambda} = 0 \Leftrightarrow \partial_\mu \rho_\Lambda = 0 \label{on_shell_lambda}
\end{equation}
\begin{equation}
    \frac{\delta S}{\delta \rho_\Lambda} = \frac{\delta S_0}{\delta \rho_\Lambda} + \frac{\delta S_U}{\delta \rho_\Lambda} = 0 \Leftrightarrow \sqrt{-g} = \partial_\mu \mathcal{T}^\mu .\label{unimodular_const}
\end{equation}
Equation \eqref{on_shell_lambda} shows how the addition of $S_U$ to the base action $S_0$ converts the cosmological constant from a fixed parameter to an integration constant obeying this equation of motion. This procedure of demoting arbitrarily fixed parameters to on-shell-only constants has been called “deconstantization”. Equation \eqref{unimodular_const} can be interpreted as a time equation for the vector density, providing a relationship between unimodular time and 4-volume ” (see~~\cite{unimod1,unimod,UnimodLee1,alan,daughton,sorkin1,sorkin2} for extensive discussions).

%~\cite{Misner_4_volume}. 

Considering a $3+1$ split of the action, to assess the Hamiltonian structure of the theory, it is found that the gauge-invariant zero-mode of $\mathcal{T}^0$,
\begin{equation}
    T \equiv \int{d^3 x \: \mathcal{T}^0} \label{time}
\end{equation}
is the canonical conjugate of $\rho_\Lambda$. It has been variously argued ~\cite{unimod1,unimod,UnimodLee1,alan,daughton,sorkin1,sorkin2,gielen, gielen1} that $T$ provides an excellent physical definition of time. In light of equation \eqref{unimodular_const}, we see that the unimodular time is just
\begin{equation}\label{TLambda}
    T_\Lambda = - \int{d^4 x \: \sqrt{-g}},
\end{equation}
i.e. 4-volume time.

Extending these ideas to constants other than $\Lambda$, the term $S_U$ in action \eqref{unimodular_1} can be written as ~\cite{JoaoPaper, JoaoLetter}:
\begin{equation}
    S = S_0 - \int{ d^4 x \: \bm{\alpha} \cdot \partial_\mu \mathcal{T}^\mu_{\bm{\alpha}}}
\end{equation}
where $\bm{\alpha}$ is a $D$-dimensional vector containing a selection of the natural constants that appear in the base action $S_0$, and where the dot denotes the Euclidean inner product for a $D$-dimensional space. Similarly to the original unimodular case, the vector density $\mathcal{T}^\mu_{\bm{\alpha}}$ provides the relational times $T_{\bm{\alpha}}$, dual to the on-shell constants $\bm{\alpha}$, since after an integration by parts the action reads:
\begin{equation}
    S=S_0+\int dt\, \dot {\bm\alpha} {\bm T}_{\bm\alpha}
\end{equation}
with ${\bm T}_{\bm\alpha}$ defined as in (\ref{time}). 
%In principle, the $\bm{\alpha}$ vector could contain any constant appearing in the base action $S_0$, which will be the Brans-Dicke action in this investigation. Therefore, t

To implement evolving physical laws, and avoid introducing second class constraints, we create a new 
%D-dimensinal 
vector of natural constants called $\bm{\beta}$ disjoint from $\bm\alpha$. These constants are then postulated ~\cite{evolution} to be dependent on the times provided by the $\bm{\alpha}$ ones, such that 
\begin{equation}
    \bm{\beta} = \bm{\beta} (\bm{T_{\alpha}})
\end{equation}
Thus, as in ~\cite{evolution},  we separate the natural constants in two groups, one providing constants dual to evolution times and the other one giving the actual constants undergoing the evolution. As we will see in this work, the membership to one set or the other is neither strict nor given \textit{a priori}, as we will consider several theoretical scenarios with different choices. As a starting point we choose as a parameterization for constant-clocks:
\begin{equation}\label{alphadef}
    \boldsymbol{\alpha} = (\alpha_1, \alpha_2,\alpha_3) = \left(\rho_\Lambda, \frac{3 c_P^2}{8\pi G_P},\frac{G_M}{G_P}\right), 
\end{equation}
where $\alpha_1$ is the $\Lambda$-vacuum energy density $\rho_{\Lambda}$, $\alpha_2$ is the reduced Planck mass squared (or Planck energy) $M^2_{\text{P}}$, and $\alpha_3$ may be invoked should we separate the gravitational coupling $G_M$ to the $G_P$ appearing in the Planck mass~\cite{vikman,vikman1,vikman2}. Beside $T_1\equiv T_\Lambda$, we have
\begin{eqnarray}
    T_2\equiv T_R&=&\frac{1}{6}\int d^4 x \sqrt{-g} R\label{TRicci}\\
     T_3\equiv T_N&=&\int d^4 x \sqrt{-g} {\cal L}\label{TMatter}
_M\end{eqnarray}
% *** INCLUDE HERE?
%[SIGNS CHECK; REFER FORWARD]
i.e.: Ricci and Newton times.
%(see~\cite{pad,pad1,lomb,vikman,vikman1} in addition to~\cite{unimod1,unimod,UnimodLee1,alan,daughton,sorkin1,sorkin2,Bombelli,UnimodLee2}). 
$T_\Lambda$ and $T_R$ are nothing but the ``fluxes'' used in the sequester model~\cite{JoaoPaper,JoaoLetter,pad,pad1}.

We will explore many possibilities for $\bm\beta$, including the matter and gravity speed of light $c_m$ and $c_g$, for example:
\begin{equation}
    \bm\beta=(c_m^2,c_g^2)
\end{equation}
as well as other choices and combinations. In particular we will be including the Brans-Dicke coupling $\omega$ and scalar $\phi$ into this game.

\subsection{Brans-Dicke theory of gravity}
The Brans-Dicke theory of gravity is intimately connected with the idea of varying constants and Mach's principle. In fact, Brans and Dicke's seminal paper ~\cite{BD} was motivated to develop a theory of gravity that, unlike general relativity ~\cite{mach_conflict}, would fully incorporate Mach's principle.
According to one formulation of Mach's principle, the geometry of space-time, and hence the inertial properties of every infinitesimal test particle, are determined by the mass-energy distribution throughout the universe. In other words, the local physics of a system is influenced by the global structure of the universe around it, suggesting non-locality.

Following ~\cite{BD}, it is possible to show that 
\begin{equation}
    G \sim \frac{R}{M}
\end{equation}
where $G$ is Newton's gravitational constant, $R$ is radius of the observable universe, $M$ is the total mass of the observable universe and where we have set the $c^2$ factor to one. This relationship confirms Mach's principle: the inertia experienced by an accelerated observer relative to distant matter is equivalent to a gravitational force acting on a fixed observer due to distant accelerated matter~\cite{BD}.

On the other hand, if the total matter in the universe is assumed to be constant (as one should hope), the above ratio can be constant only if $R$ is fixed. However, form observations ~\cite{Supernova}, the universe is expanding, so that $R$ is a varying parameter. Therefore, it is straightforwardly concluded that also $G$ will vary over some evolution parameter $t$.

This results provides the essential link between Mach's principle and a varying gravitational constant, leading to the Brans-Dicke's modified gravity action. Starting from the Einstein-Hilbert (EH) action, we include a coordinate independent scalar filed to model a varying gravitational constant, defined as
\begin{equation}
    \phi : = \frac{1}{G}
\end{equation}

Then, multiplying the EH action by $G^{-1}$ and including a Lagrangian density for the scalar field we obtain the Brans-Dicke gravity action~\cite{BD}
\begin{equation}
    S_{BD} = \frac{c^4}{16 \pi}\int{d^4 x \: \sqrt{-g} \: \biggl [\phi R - \frac{\omega}{\phi} g^{\mu \nu} \: \partial_\mu \phi  \: \partial_\nu \phi\biggl]} ,
   % + S_M 
    \label{BD_action}
\end{equation}
where the scalar field's kinetic term derives from minimal dimension consistency arguments and that $\phi$ is expected to obey a second order wave-like equation.
We must add to this a generic matter action, $S_M$, assumed not to contain $\phi$. This enforces conservation of the energy-momentum tensor of matter (and geodesic motion for point particles). 

By including the factor $\phi^{-1}$ in the kinetic term,
the parameter $\omega$ is rendered a dimensionless coupling constant, which can \textit{a priori} take any value. Modulo some exceptions, we recover Einstein’s theory in the limit where $\omega \rightarrow \infty$. The rest of the Lagrangian density for $\phi$ has the form of a kinetic term. 
%This is actually the case, as this term will provide the scalar's dynamical evolution. 
%THIS IS WRONG, INDEED WE DO NOT NEED IT LATER
%Furthermore, the kinetic term’s introduction assures full diffeomorphism’s invariance of the action and the conservation of the energy-momentum tensor via $\nabla_\mu T^{\mu \nu}$ [64]. 
It is also possible to include a general potential $V(\phi)$ ~\cite{BD_potential}, 
%thus arriving at an action with full kinetic and potential terms for $\phi$, 
but, in this paper, we will neglect it.

To conclude, the Brans-Dicke theory has two main advantages: firstly, it directly incorporates a varying gravitational constant into the theory's action via Mach's principle. Secondly, it introduces an additional parameter $\omega$ in comparison to classical GR, which will be used to enlarge the set of constants forming the $\bm{\alpha}$ or the $\bm{\beta}$ vectors. Notably, the value of $\omega$ is crucial to differentiate classical GR from Brans-Dicke, which, on the Solar System scale, has been mesured to be of order $\omega > 4 \times 10^4$ ~\cite{cassini}, making the two theories equivalent. 

\section{Reduction to Homogeneity and Isotropy}\label{reduct}
For simplicity, for the rest of
this paper we will assume homogeneity and isotropy (and use the terminology of  minisuperspace or MSS). Hence we start by reducing the full Brans-Dicke action (\ref{BD_action}) to homogeneity and isotropy, resulting in: 
\begin{equation}
     S_{BD} = V_c \int{dt \: \biggl[\phi \dot{b}a^2 + \dot{\phi}\pi_\phi + \phi Na (b^2 + kc_g^2) -\frac{1}{4}\frac{\phi \pi_\phi^2 }{\omega a^3}N\biggl]} \label{BD_MSS}
\end{equation}
where $a$ is the scale factor, $b$ is the minisuperpace connection variable, satisfying $b = \frac{\dot{a}}{N}$ if the torsion-free condition is assumed, $N$ is the lapse function and $V_c$ is comoving volume. Furthermore, $c_g$ is the gravity speed of light, originating from the gravity metric $g_{\mu \nu}$, which should be differentiated from $c_P$ (appearing in the Planck scale) and $c_m$ (the speed of light in the matter metric), as pointed out in  ~\cite{EllisVSL}.
The first and third terms are those found in standard GR, multiplied by $\phi$. The last term is the Brans-Dicke Hamiltonian, while the second term is the canonical pair of the scalar field and its conjugate momentum $\pi_\phi$, 
%[defined as] if you wish; in practice this comes out of an EOM in the ham framework. 
\begin{equation}
    \pi_\phi := \frac{\delta S_{BD}}{\delta \dot{\phi}} = 2 \omega \frac{\dot{\phi}}{\phi Nc_\phi}a^3 \label{def_pi_phi}
\end{equation}
%Importantly, 
We highlight the factor of $c_\phi$ in the denominator, this is the speed of propagation for the Brans-Dicke field. 
%while we will set it to one in some cases, it will be essential in others as an additional constant that may be varying. 
While we will set it to a constant in this paper it can be used to produced further models, as we will do in~\cite{cphi}

Similarly reducing the matter base action $S_M$: 
\begin{equation}
    S_M = V_c \int{dt \: \alpha_3 \biggl[\dot{m}_i \psi_i - Na^3 (\rho_\Lambda + \rho_M)\biggl]} \label{S_M}
\end{equation}
and the unimodular-like action $S_U$:
\begin{equation}\label{SUred}
    S_U = V_c \int{dt \: \bm{\dot{\alpha}} \cdot \bm{T} \:}
\end{equation}
where $\alpha_3$ is a matter constant defined as $\alpha_3 = G_N/G_P$.

In equation \eqref{S_M} above, we allow for a matter action with multi-fluids, such that $\rho_M = \sum_i \rho_i$, where
\begin{equation}
    \rho_i = \frac{m_i}{a^{3(1+w_i)}} \label{rho_def}
\end{equation}
with $w_i$ being the usual fluid's equation of state, $m_i$ a constant of motion dual to the variable $\psi_i$ as defined in the Lagrangian formulation of perfect fluid ~\cite{brown}, reduced to MSS as ~\cite{gielen, gielen1, twot}.

Hence, noting that the full action is given by $S = S_0 + S_U = S_{BD} + S_M + S_U$, we have the total reduced Hamiltonian: 
\begin{equation}
    H = Na \biggl[-\phi (b^2 + kc_g^2) + \alpha_3 (\rho_\Lambda + \rho_M)a^2 + \frac{1}{4}\frac{\phi \pi_\phi^2}{\omega a^4} \biggl] \label{BD_H}.
\end{equation}
Crucially, we must bear in mind that the canonical variable dual to $b$ (needed for evaluating Poisson brackets) is $A^2 = \phi a^2$ and not $a^2$ as usual, because $\phi$ multiplies the canonical term in action \eqref{BD_MSS}.

\section{The central forks in the formalism}\label{forks}

From this point onward, the theory branches off as a result of two separate forks. First, we can differentiate the cases where we consider a dynamical scalar field from the ones where we assume a non-dynamical one (i.e., when $\omega = 0$). Second, we can consider the cases where the scalar field $\phi$ in the first canonical term of action \eqref{BD_MSS} is kept constant, such that $\phi \rightarrow \phi_0$, while the one in the Hamiltonian is left free to change. %On the other hand, we also analyse scenarios were $\phi$ is a varying parameter in all the terms of action \eqref{BD_MSS}, following the spirit of the original Brans-Dicke theory. Importantly, in the former case, we fully implement local Lorentz invariance violation (LLIV), as prescribed by the Horava-Lifshitz theory [REF].

\subsection{Non-Dynamical Brans-Dicke}

We first set $\omega = 0$, and present the local Lorentz invariance violating case, where $\phi \rightarrow \phi_0$ in the canonical term, for a non-dynamical Brans-Dicke action reduced to MSS. For $\omega=0$ We can neglect the case where $\phi$  varies in both the canonical term and the Hamiltonian, as it is equivalent to the $\alpha_2$ theory presented in~\cite{evolution}. Therefore, setting $\omega = 0$ in \eqref{BD_action} and reducing it to MSS we have  
\begin{equation}
    S_{\omega = 0} = V_c \int{dt \:  \phi_0 \biggl[\dot{b}a^2 + Na (b^2 +kc_g^2)\biggl]}
\end{equation}
which, combined with $S_M$, gives the non-dynamical Brans-Dicke Hamiltonian:
\begin{equation}
    H = Na \biggl[-\phi (b^2 + kc_g^2) + \alpha_3 (\rho_\Lambda + \rho_M)a^2 \biggl].
\end{equation}
Setting $H \overset{!}{=} 0$, we have the Hamilton's constraint,
\begin{equation}
    b^2 + kc_g^2 = \frac{\alpha_3 \rho a^2}{\phi} \label{Hamilton_const_1}
\end{equation}
while, using Poisson's brackets, we arrive at the equations of motion for the canonical pair $(b, a^2)$. Crucially, since $\phi_0$ is fixed in the canonical term, we can use $a^2$ and $A^2=a^2\phi_0$ interchangeably when evaluating the brackets (see~\cite{evolution} for the problems that arise otherwise). Therefore, we have
\begin{equation}
    \dot{a} = \{a, H \} = \frac{\phi}{\phi_0}Nb \label{a_eqn_non_dyn}
\end{equation}
\begin{equation}
    \dot{b} = \{b, H \} = -\frac{\alpha_3}{2 \phi_0} (\rho +3p)Na
\end{equation}
where we have used equation \eqref{rho_def} when deriving the $\dot{b}$ equation of motion. Finally, we wish to obtain a cosmological conservation equation to gain insights on energy source/sinks terms due to varying constants in this theory. Therefore, dotting equation \eqref{Hamilton_const_1} and substituting the equations of motion, we obtain 
\begin{equation}
    \dot{\rho} + 3\frac{\dot{a}}{a}(\rho + p) = -\frac{\dot{\alpha_3}}{\alpha_3}\rho + \frac{\dot{\phi}}{\phi}\rho + \frac{k \phi }{\alpha_3 a^2}\frac{dc_g^2}{dt} \label{cons_eqn}
\end{equation}
This is the most general conservation equation assuming all the constants are varying, and will be the heart of specific constants' considerations in the next section. Remarkably, this scenario has a $\phi$ scalar field behaving like an $\alpha_3$ theory due to LLIV. Most of what follows regards the behaviour of different theories regarding this non-conservation equation, which is central to early Universe modelling. 

\subsection{Dynamical Brans-Dicke}

We next consider the fully dynamical Brans-Dicke action. In this scenario, both the LIVBD and the Lorentz invariant cases generate original results, so we will consider both.

Considering first the LIVBD case, we use Hamiltonian \eqref{BD_H}, leading to the new Hamilton's constraint: 
\begin{equation}
    b^2 + kc_g^2 = \frac{\alpha_3 \rho a^2}{\phi} + \frac{1}{4}\frac{\pi_\phi^2}{\omega a^4} \label{H_const_full}
\end{equation}
where the additional factor is arising from the Brans-Dicke kinetic term. The $\dot{a}$ equation is the same one as \eqref{a_eqn_non_dyn}, while the $\dot{b}$ equation reads 
\begin{equation}
    \dot{b} = \{b, H\} = -\frac{\alpha_3}{2 \phi_0}(\rho + 3p)Na -\frac{1}{2}\frac{\phi \pi_\phi^2}{\phi_0 \omega a^5}N.
\end{equation}
Furthermore, we have the equation of motion for the conjugate momentum $\pi_\phi$:
\begin{equation}
    \dot{\pi}_\phi = \{\pi_\phi, H\} = \frac{\alpha_3 \rho}{\phi}Na^3 \label{pi_dot_short}
\end{equation}
where we have used constraint \eqref{H_const_full}. Finally, dotting the constraint and using the equations of motion, we obtain 
\begin{equation}
       \dot{\rho} + 3\frac{\dot{a}}{a}(\rho +p) = -\frac{\dot{\alpha}_3}{\alpha_3}\rho +\frac{1}{4}\frac{\phi \pi_\phi^2}{\alpha_3 \omega^2 a^6}\dot{\omega} + \frac{k \phi}{\alpha_3 a^2}\frac{dc_g^2}{dt} \label{cons_eqn_full}
\end{equation}

In the case where we do not break Lorentz invariance ($\phi$ is the same in the canonical pairs and in the rest of action), the Hamiltonian is the same one as \eqref{BD_H}, as well as the constraint, but the equations of motion differ. Crucially, when evaluating Poisson's brackets, $A^2=\phi a^2$ must be used since $\phi$ in the canonical term is varying, as already pointed out in ~\cite{evolution}. This leads to 
\begin{equation}
    \dot{a} + \frac{\dot{\phi}}{2\phi}a = Nb \label{full_a}
\end{equation}
\begin{equation}
    \dot{b} = -\frac{\alpha_3}{2 \phi}(\rho + 3p)Na -\frac{1}{2}\frac{\pi_\phi^2}{ \omega a^5}N \label{full_b}.
\end{equation}
Additionally, evaluating the Poisson's brackets for the conjugate momentum $\pi_\phi$, we obtain 
\begin{equation}
    \dot{\pi}_\phi = -Na \phi \biggl[\biggl(\frac{-\rho + 3p}{2}\biggl)\frac{\alpha_3 a^2}{\phi^2} + \frac{1}{2}\frac{\pi_\phi^2}{\omega a^4}\frac{1}{\phi}\biggl] \label{full_pi}
\end{equation}
Dotting the Hamilton's constraint and combining equations \eqref{full_a}, \eqref{full_b}, \eqref{full_pi} and using relation \eqref{def_pi_phi} for $\dot{\phi}$, we arrive at the general conservation equation, which takes the same form as \eqref{cons_eqn_full}.

\section{A selection of interesting theories}\label{selection}
At this point the number of possibilities arising from these forks and choices for $\bm\alpha$ and $\bm\beta$ lead to an ungainly proliferation of possibilities. Having sieved through them~\cite{Paolothesis}, we report here on those that are most interesting due to their simplicity and/or promising properties.

\subsection{Brans-Dicke parameter depending on unimodular time} \label{main_result}

We begin with the simplest theory: one where the parameter generating the evolution's time is $\bm{\alpha} = \rho_\Lambda$, and $\bm{\beta} = \omega$. In this setup, the unimodular time describes the evolution of the Brans-Dicke parameter, with $\omega = \omega (T_\Lambda)$. It turns out that the LIV and non-LIV BD cases lead to the same results so, for simplicity, we present the latter only. Thus, besides the constraint and equations  \eqref{full_a}, \eqref{full_b} and \eqref{full_pi} we have two additional Hamilton's equations:
\begin{equation}
    \dot{\rho}_\Lambda = \{\rho_{\Lambda}, H \} = \frac{\partial H}{\partial T_{\Lambda}} = - N\frac{1}{4}\frac{\phi \pi_\phi^2}{\omega^2 a^3} \frac{\partial \omega}{\partial T_{\Lambda}} \label{varying_lmdb}
\end{equation}
\begin{equation}
    \dot{T}_\Lambda = \{T_{\Lambda}, H \} = - \frac{\partial H}{\partial \rho_{\Lambda}} = - \alpha_3 Na^3. \label{uni_time}
\end{equation}
The first equation gives the time evolution of the cosmological constant, providing the source term in equation \eqref{cons_eqn_full}, where we keep both $\alpha_3$ and $c_g$ fixed. Combining equation \eqref{varying_lmdb} with \eqref{uni_time}, we immediately see that the source term given by $\dot{\omega}$ goes fully into the vacuum energy, returning the usual conservation equation.

Additionally, equation \eqref{uni_time} states that $T_\Lambda$ is the 4-volume, if we assume $\alpha_3 = 1$. Combining equations \eqref{varying_lmdb} and \eqref{uni_time} and using the definition of $\pi_\phi$ from \eqref{def_pi_phi} in the gauge where $N = 1$, we obtain 
\begin{equation}\label{dotLdoto}
    \dot{\rho}_\Lambda = \frac{\dot{\phi}^2}{\phi}\dot{\omega}.
\end{equation}
%where, for simplicity, we have assumed that $\alpha_3 = 1$. 
This is not ideal. 
But it would be too good to be true for such a simple scenario to generate a low vacuum energy out of driving $\omega$ to infinity (and so the theory to the GR limit). The connection is potentially there, but certainly not in such a simple theory. Specifically the sign in \eqref{dotLdoto} is wrong; in addition we should have a $\rho_\Lambda$ factor on its right hand side. This theory is simple, but it has the wrong phenomenological properties.

\subsection{Fixed $\phi_0$ with Ricci and Newton clocks} \label{Ricci_N}
Next, we consider a LIVBD scenario (with a fixed $\phi_0$ in the canonical pair term) where both $\phi$ and $\alpha_3$ produce clocks, and  $c_m^2 = c_m^2 (T_\phi, T_N)$. Here, $T_\phi$ is the Ricci time dual to $\phi$, while $T_N$ is the ``Newton'' time dual to $\alpha_3$.

We start with a clarification. By including $\phi$ in $\bm{\alpha}$, we could potentially produce a  duplication of terms, and this has to be avoided. A naive combination of \eqref{BD_MSS} and \eqref{SUred}
would produce:
\begin{equation}
    S = V_c \int{dt \: \biggl[ \phi \dot{b}a^2 + \dot{\phi}\pi_\phi - H} \biggl] + \: V_c \int{dt \: \dot{\phi}T_\phi}
\end{equation}
where we immediately notice a duplication of canonical terms, leading to contradictions unless we identify $T_\phi = \pi_\phi$ and keep only one term.

Having made this clear, we now investigate the violations of energy conservation in this theory. 
The conservation equation we should start from is the full \eqref{cons_eqn}, since all its three source terms contribute. In addition we have Hamilton's equations:
\begin{eqnarray}
    \dot{\phi} =\{\phi, H\} &=& - \phi k Na \frac{\partial c_g^2}{\partial T_\phi} + \alpha_3 Na^3 \frac{\partial \rho}{\partial c_m^2}\frac{\partial c_m^2}{\partial T_\phi} \label{phi_dot}\\
      \dot{\alpha}_3 = \{\alpha_3, H\} &=& - \phi k Na \frac{\partial c_g^2}{\partial T_N} + \alpha_3 Na^3 \frac{\partial \rho}{\partial c_m^2}\frac{\partial c_m^2}{\partial T_N} \label{alpha_3_dot}
\end{eqnarray}
where the Ricci time $T_\phi$ should be understood as the canonical momentum $\pi_\phi$, as defined in equation \eqref{def_pi_phi}. The equation of motion for the Newton time reads
\begin{equation}
    \dot{T}_N = \{T_N, H \} = -\frac{\partial H}{\partial \alpha_3} = -\rho Na^3,
\end{equation}
while the $\dot{T}_\phi$ equation is given entirely by the equation of motion for $\dot{\pi}_\phi$, i.e., by equation \eqref{pi_dot_short}. 
Expanding the $c_g^2$ source term in \eqref{cons_eqn} in terms of the Ricci and Newton times and using equations \eqref{phi_dot} and \eqref{alpha_3_dot}, we finally obtain:
\begin{equation}
     \dot{\rho} + 3 \frac{\dot{a}}{a} (\rho + p) = \rho Na^3\frac{\partial \rho}{\partial c_m^2}\biggl[\frac{\alpha_3}{\phi}  \frac{\partial c_m^2}{\partial T_\phi} - \frac{\partial c_m^2}{\partial T_N} \biggl]
\end{equation}
We note that there is indeed a source term, giving net energy violation. This happens because of the dependence of $c_m^2$ on the Ricci and Newton times and is part of a larger pattern where matter parameters depending on gravitational clocks lead to energy violation in the conservation equation. It turns out this is the only case where the LLIV and non-LIV scenario give different answers in this respect. The term in the Ricci time contains does not contain the pressure, if there is Lorentz invariance violation; it does otherwise, as proved in the conservation equation of section VI B in~\cite{evolution}.

\subsection{Brans-Dicke varying speed of light (VSL)} \label{c_g_section}

The following case is perhaps the most interesting, a priori, for phenomenology. It consists of 
the usual scenario where the speed of light $c_g$ is assumed to be varying; however, unlike in previous work~\cite{AM,MoffatVSL,VSLreview,evolution}, $c_g$ is made to depend on the Brans-Dicke time, so that $c_g^2 = c_g^2 (T_\omega)$. Therefore we have a scenario of a gravitational parameter depending on clock also given by a gravitational parameter.

Since $c_g$ is the only constant assumed to be varying in this scenario, the general conservation equation \eqref{cons_eqn_full} takes the form 
\begin{equation}
    \dot{\rho} + 3\frac{\dot{a}}{a}(\rho +p) = \frac{k \phi}{\alpha_3 a^2}\frac{dc_g^2}{dt}.
\end{equation}
However, given the additional Hamilton's equation for $\omega$, the conservation equation requires also the $\dot{\omega}$ term present in \eqref{cons_eqn_full}. Therefore, computing the Hamilton's equations for the constant and its canonical time, we obtain 
\begin{equation}
    \dot{\omega} = \{\omega, H \} = \frac{\partial H}{\partial T_\omega} = -k \phi Na \frac{\partial c_g^2}{\partial T_\omega} \label{omega_eqn}
\end{equation}
\begin{equation}
    \dot{T}_\omega = \{T_\omega, H \} = -\frac{\partial H}{\partial \omega} = \frac{1}{4}\frac{\phi \pi_\phi^2}{\omega^2 a^3}N \label{BD_time}
\end{equation}
First, we see that substituting \eqref{omega_eqn} into the $\omega$ term of the conservation equation and expanding the time dependence of $c_g$ in terms of $T_\omega$ we recover full energy conservation. This is an expected results since, as it was pointed out in ~\cite{evolution}, gravitational parameters depending on gravitational clocks do not lead any net energy violation, not producing any matter.

Second, using equation \eqref{omega_eqn} combined with the expression for the evolution of $T_\omega$, we arrive at a relation between $\omega$ and $c_g$:
\begin{equation}
    \dot{\omega} = -\frac{k}{a^2}\frac{\phi^2}{\dot{\phi}^2}\dot{c}_g^2
\end{equation}
where we have used the gauge $N = 1$. Clearly, if $c_g$ underwent a phase transition from its high value in the early universe to the value we currently observe as postulated in ~\cite{AM}, then this equation suggests a concomitant increase in $\omega$, if $k=1$. This mechanism explains the current bounds on $\omega$, and why gravity is so close to GR, even if $\omega$ was originally small. 

%However, this result depends crucially on the value of the spatial curvature $k$. In fact, a universe with $k = 0,-1$ would certainly be problematic, as much as the scenario in \ref{main_result}. 

\subsection{Matter speed of light with Brans-Dicke time}
In the previous case the gravity speed of light varied with the time dual to $\omega$, but what if the matter speed of light is different, and it is the relevant parameter in $\bm\beta$? Here 
we present the case where $\bm{\alpha} = \omega$ and $\bm{\beta} = c_m^2$, with $c_m^2 = c_m^2 (T_\omega)$.

We consider the non-LIVBD case.  Then, the conservation equation \eqref{cons_eqn_full} for a varying $c_m^2$, including the source term given by $c_g^2$, is 
\begin{equation}
    \dot{\rho} + 3\frac{\dot{a}}{a}(\rho +p) =
    \frac{1}{4}\frac{\phi \pi_\phi^2}{\alpha_3 \omega^2 a^6}\dot{\omega} + \frac{k \phi}{\alpha_3 a^2}\frac{dc_g^2}{dt} \label{cons_omega_cm}
\end{equation}
where we have excluded a varying $\alpha_3$. Additionally, we have the two Hamilton's equations, reading 
\begin{equation}
    \dot{\omega} = \{\omega, H\} = \frac{\partial H}{\partial T_\omega} = - k \phi Na \frac{\partial c_g^2}{\partial T_\omega} +\alpha_3 Na^3 \frac{\partial \rho}{\partial c_m^2}\frac{\partial c_m^2}{\partial T_\omega} \label{omega_add}
\end{equation}
\begin{equation}
    \dot{T}_\omega = \{T_\omega, H \} = -\frac{\partial H}{\partial \omega} = \frac{1}{4}\frac{\phi \pi_\phi^2}{\omega^2 a^3}N \label{BD_time_2}
\end{equation}
Note that the $\omega$ equation differs from \eqref{omega_eqn} because of the additional term given by the $\rho$ derivatives. Central to this fact is the specification of the chemical potentials $\rho_i = \rho_i (n_i, c_m)$ for each fluid species $i$, where $n_i$ is the volume density of a conserved particle number. The time equation is the same as \eqref{BD_time} as we might expect, since the Hamiltonian does not show any other explicit dependencies on $\omega$. Substituting \eqref{omega_add} in \eqref{cons_omega_cm} and using equation \eqref{BD_time_2} explicitly in the $\dot{c}_g^2$ term, we obtain
\begin{equation}
    \dot{\rho} + 3\frac{\dot{a}}{a}(\rho +p) = \frac{\dot{\phi}^2}{\phi}a^3 \frac{\partial \rho}{\partial c_m^2}\frac{\partial c_m^2}{\partial T_\omega}
\end{equation}
We see that, for a matter parameter depending on a gravitational clock, there is net energy violation, given by the variation of the clock. This result is not unexpected, as it is part of the larger pattern already identified in ~\cite{evolution}, which is now confirmed to hold here for Brans-Dicke theories.

Furthermore, we can go beyond this result, deriving a relationship between the Brans-Dicke parameter and the matter speed of light $c_m^2$. First, we take equation \eqref{omega_add}, setting the $c_g^2$ source term to zero since we want a relationship for $c_m^2$ only. Then, using the Brans-Dicke time expression as well as equation \eqref{def_pi_phi} we obtain 
\begin{equation}
    \dot{\omega} = \alpha_3 \frac{\phi}{\dot{\phi}^2}\frac{\partial \rho}{\partial c_m^2}\dot{c}_m^2
\end{equation}
Unlike in Section \ref{c_g_section}, the matter speed of light grows with the same sign as the Brans-Dicke parameter. Also, there is no dependence on the curvature $k$, as well as on the scale factor. Face value this can be bad, but we hope to return to these models in a future implication. They certainly imply an interesting role for $c_m$.

\subsection{Ricci and Brans-Dicke times}

We finally generalise the previous result, allowing $c_m^2$ to depend on both the Ricci and Brans-Dicke times. This is the case where $c_m^2 = c_m^2 (T_\phi, T_\omega)$, that is, $\bm{\alpha} = (\phi, \omega)$ and $\bm{\beta} = c_m^2$. Unlike Section \ref{Ricci_N}, the conservation equation cannot be straightforwardly used, as the $\dot{\phi}$ equation is different from when $\phi$ is not assumed to be varying. This occurs because of the extra terms in $\dot{\phi}$ given by the dependency of the Hamiltonian on multiple relational times. Thus, we first obtain the equation of motion for $\phi$:
\begin{equation}
    \dot{\phi} = \{\phi, H\} = \frac{1}{2}\frac{\phi \pi_\phi N}{\omega a^3} +\alpha_3 Na^3 \frac{\partial \rho}{\partial c_m^2}\frac{\partial c_m^2}{\partial T_\phi} -\phi Na k \frac{\partial c_g^2}{\partial T_\phi} \label{phi_dot_long},
\end{equation}
whilst the equations of motion for $\omega$ and its time are the same ones as equations \eqref{omega_add} and \eqref{BD_time}. Furthermore, the Ricci time equation is the same as the equation of motion for the canonical momentum, i.e., as equation \eqref{full_pi}.

We can now dot the constraint \eqref{H_const_full} to obtain a general expression, where we then use equation \eqref{phi_dot_long} in the $\dot{\phi}$ terms. This leads to a conservation equation with a profusion of terms, which after some tedious algebra simplifies to 
\begin{multline}
     \dot{\rho} +3\frac{\dot{a}}{a}(\rho + p) = \frac{\partial \rho}{\partial c_m^2} \biggl[\frac{\alpha_3 Na^3}{\phi}\frac{(\rho -3p)}{2}\frac{\partial c_m^2}{\partial T_\phi} \\ \\ -\frac{1}{2}\frac{\pi_\phi^2 N}{\omega a^3}\frac{\partial c_m^2}{\partial T_\phi} +\frac{1}{4}\frac{\phi \pi_\phi^2 N}{\omega^2 a^3}\frac{\partial c_m^2}{\partial T_\omega}\biggl]
\end{multline}
The first source term is the same as in the  $c_m^2 = c_m^2 (T_R, T_N)$ scenario, already studied in  ~\cite{evolution}, where both $\alpha_2$ and $\alpha_3$ are assumed to be varying. The second and third source terms are new, and unique to Brans-Dicke models. It would be interesting to build models of the early Universe based on them, to see how they interact with the usual cosmological problems.

\section{Conclusions}

In this paper, we investigated a hybrid between Brans-Dicke theory and unimodular-like varying constants, focusing on two focal points: cosmological energy conservation violations and the correlated variation of multiple constants. These two questions are intimately connected to each other and to the ideas of unimodular gravity and Mach's principle. Motivated by the latter and inspired by earlier work~\cite{evolution}, we extended the formalism beyond General Relativity to Brans-Dicke gravity. With plenty of possibilities and several applications, some results merely confirmed and extended previous work, while others were new and set the stage to further applications  (such as within the sequester model~\cite{pad,pad1}).

We found that the energy conservation violation pattern previously found for matter and gravity clocks and constants also holds for the Brans-Dicke theory. We verified that energy is produced only in two possible scenarios: a matter parameter varying in terms of a gravity clock or a gravity parameter evolving with a matter clock. We confirmed that this pattern still holds in Brans-Dicke theories. 
%as resulting from the field equations' non-linearity. 

Regarding joint variations of different constant, multiple forks in the formalism were explored, considering the combinations of dynamical and non-dynamical Brans-Dicke, with and without Lorentz invariance violation, and several choices for evolution potentials $\bm\beta (\bm T_{\bm\alpha})$. A comprehensive account can be found in~\cite{Paolothesis}. It turned out that LIV vs non-LIV made little difference to the outcomes. By far the most interesting scenario, face-value, is that with $c_g^2 (T_\omega)$, where the speed of gravity depends on a clock dual to the Brans-Dicke coupling $\omega$. For a generic potential $c_g^2 (T_\omega)$ we seem to find a relation between solutions to the horizon, homogeneity and flatness problem and the reason why $\omega$ must currently be so large, and gravity so close to GR. But the other scenarios presented in Section~\ref{selection}, although clearly flawed, also have potential, assuming that they can be inserted into suitably modified theories.   

One must stress that in the model building game for the early Universe, even within the evolution game, there are not that many gravitational parameters to play with. Beside the speed of gravity one is confined to Newton's constant, which, if one wishes, may be split in its separate dual roles of Planck mass and gravitational coupling~\cite{vikman,vikman1}. The addition of $\omega$ could be a game changer in this respect. This paper established the background for further work.

\section{Acknowledgments}
We thank Lorenzo Signore for helpful discussions related to this paper. The work of JM was partly supported by the STFC Consolidated Grants ST/T000791/1 and ST/X00575/1.


\begin{thebibliography}{0}%
\makeatletter
\providecommand \@ifxundefined [1]{%
 \@ifx{#1\undefined}
}%
\providecommand \@ifnum [1]{%
 \ifnum #1\expandafter \@firstoftwo
 \else \expandafter \@secondoftwo
 \fi
}%
\providecommand \@ifx [1]{%
 \ifx #1\expandafter \@firstoftwo
 \else \expandafter \@secondoftwo
 \fi
}%
\providecommand \natexlab [1]{#1}%
\providecommand \enquote  [1]{``#1''}%
\providecommand \bibnamefont  [1]{#1}%
\providecommand \bibfnamefont [1]{#1}%
\providecommand \citenamefont [1]{#1}%
\providecommand \href@noop [0]{\@secondoftwo}%
\providecommand \href [0]{\begingroup \@sanitize@url \@href}%
\providecommand \@href[1]{\@@startlink{#1}\@@href}%
\providecommand \@@href[1]{\endgroup#1\@@endlink}%
\providecommand \@sanitize@url [0]{\catcode `\\12\catcode `\$12\catcode
  `\&12\catcode `\#12\catcode `\^12\catcode `\_12\catcode `\%12\relax}%
\providecommand \@@startlink[1]{}%
\providecommand \@@endlink[0]{}%
\providecommand \url  [0]{\begingroup\@sanitize@url \@url }%
\providecommand \@url [1]{\endgroup\@href {#1}{\urlprefix }}%
\providecommand \urlprefix  [0]{URL }%
\providecommand \Eprint [0]{\href }%
\providecommand \doibase [0]{https://doi.org/}%
\providecommand \selectlanguage [0]{\@gobble}%
\providecommand \bibinfo  [0]{\@secondoftwo}%
\providecommand \bibfield  [0]{\@secondoftwo}%
\providecommand \translation [1]{[#1]}%
\providecommand \BibitemOpen [0]{}%
\providecommand \bibitemStop [0]{}%
\providecommand \bibitemNoStop [0]{.\EOS\space}%
\providecommand \EOS [0]{\spacefactor3000\relax}%
\providecommand \BibitemShut  [1]{\csname bibitem#1\endcsname}%
\let\auto@bib@innerbib\@empty
%</preamble>
\end{thebibliography}%


\begin{thebibliography}{99}
\bibitem{dirac}
P.A.M. Dirac, Nature (London) 139 (1937) 323.

%\cite{Brans:1961sx}
\bibitem{BD}
C.~Brans and R.~H.~Dicke,
%``Mach's principle and a relativistic theory of gravitation,''
Phys. Rev. \textbf{124}, 925-935 (1961)
doi:10.1103/PhysRev.124.925
%4169 citations counted in INSPIRE as of 07 Nov 2023

\bibitem{Bek}
J.~D.~Bekenstein,
%``Fine Structure Constant: Is It Really a Constant?,''
Phys. Rev. D \textbf{25}, 1527-1539 (1982)
doi:10.1103/PhysRevD.25.1527

\bibitem{BSM}
H.~B.~Sandvik, J.~D.~Barrow and J.~Magueijo,
%``A simple cosmology with a varying fine structure constant,''
Phys. Rev. Lett. \textbf{88}, 031302 (2002)
doi:10.1103/PhysRevLett.88.031302
[arXiv:astro-ph/0107512 [astro-ph]].


\bibitem{emass}
J.~D.~Barrow and J.~Magueijo,
%``Cosmological constraints on a dynamical electron mass,''
Phys. Rev. D \textbf{72}, 043521 (2005)
doi:10.1103/PhysRevD.72.043521
[arXiv:astro-ph/0503222 [astro-ph]].

\bibitem{CovVSL}
J.~Magueijo,
%``Covariant and locally Lorentz invariant varying speed of light theories,''
Phys. Rev. D \textbf{62} (2000), 103521
doi:10.1103/PhysRevD.62.103521
[arXiv:gr-qc/0007036 [gr-qc]].

\bibitem{eorc}
J.~Magueijo, J.~D.~Barrow and H.~B.~Sandvik,
%``Is it e or is it c? Experimental tests of varying alpha,''
Phys. Lett. B \textbf{549}, 284-289 (2002)
doi:10.1016/S0370-2693(02)02928-3
[arXiv:astro-ph/0202374 [astro-ph]].

\bibitem{dagny}
D.~Kimberly and J.~Magueijo,
%``Varying alpha and the electroweak model,''
Phys. Lett. B \textbf{584}, 8-15 (2004)
doi:10.1016/j.physletb.2004.01.050
[arXiv:hep-ph/0310030 [hep-ph]].

\bibitem{unimod} M.~Henneaux and C.~Teitelboim, ``The cosmological constant and general covariance,'' {\em Phys.\ Lett.\ B} {\bf 222} (1989), 195--199.

\bibitem{unimod1}
W. G. Unruh, Phys. Rev. {\bf D40}, 1048 (1989);  K.~V.~Kucha\v{r}, ``Does an unspecified cosmological constant solve the problem of time in quantum gravity?,'' {\em Phys.\ Rev.\ D} {\bf 43} (1991), 3332--3344.


\bibitem{UnimodLee1}
L.~Smolin, ``Quantization of unimodular gravity and the cosmological constant problems,'' {\em Phys.\ Rev.\ D} {\bf 80} (2009), 084003, arXive: 0904.4841.

\bibitem{alan} A. Daughton, J. Louko, and R. D. Sorkin, ``Instantons and unitarity in quantum cosmology with fixed four-volume,'' {\em Phys.\ Rev.\ D} {\bf 58}, 084008 (1998).

\bibitem{daughton} A. Daughton, J. Louko, and R. D. Sorkin, ``Initial conditions and unitarity in unimodular quantum cosmology,'' [gr-qc/9305016].

\bibitem{sorkin1} R. D. Sorkin, ``Role of time in the sum-over-histories framework for gravity,'' {\em Int J Theor Phys} {\bf 33}, 523–534 (1994). https://doi.org/10.1007/BF00670514

\bibitem{sorkin2} R. D. Sorkin, ``Forks in the road, on the way to quantum gravity,'' {\em Int J Theor Phys} {\bf 36}, 2759–2781 (1997). https://doi.org/10.1007/BF02435709


%\bibitem[Hartle and Hawking(1983)]{HH}J.~B.~Hartle and S.~W.~Hawking,``Wave Function of the Universe,'' {\em Phys.\ Rev.\ D} \textbf{28} (1983), 2960--2975.


\bibitem{Bombelli}
L.~Bombelli, W.~E.~Couch and R.~J.~Torrence,
%``Time as space-time four volume and the Ashtekar variables,''
Phys. Rev. D \textbf{44}, 2589-2592 (1991)
doi:10.1103/PhysRevD.44.2589

\bibitem{UnimodLee2}
L.~Smolin,
%``Unimodular loop quantum gravity and the problems of time,''
Phys. Rev. D \textbf{84}, 044047 (2011)
doi:10.1103/PhysRevD.84.044047
[arXiv:1008.1759 [hep-th]].

\bibitem{pad}
N.~Kaloper and A.~Padilla,
%``Sequestering the Standard Model Vacuum Energy,''
Phys. Rev. Lett. \textbf{112}, no.9, 091304 (2014).

\bibitem{pad1}
N.~Kaloper, A.~Padilla, D.~Stefanyszyn and G.~Zahariade,
%``Manifestly Local Theory of Vacuum Energy Sequestering,''
Phys. Rev. Lett. \textbf{116}, no.5, 051302 (2016)
doi:10.1103/PhysRevLett.116.051302
[arXiv:1505.01492 [hep-th]].

\bibitem{lomb}
L.~Lombriser,
%``On the cosmological constant problem,''
Phys. Lett. B \textbf{797}, 134804 (2019).



\bibitem{weinberg}
%\bibitem{Weinberg:1988cp}
S.~Weinberg,
%``The Cosmological Constant Problem,''
Rev. Mod. Phys. \textbf{61}, 1-23 (1989).
%doi:10.1103/RevModPhys.61.1

\bibitem{padilla}
%\bibitem{Padilla:2015aaa}
A.~Padilla,
``Lectures on the Cosmological Constant Problem,''
[arXiv:1502.05296 [hep-th]].

\bibitem{evolution}
J.~Magueijo,
%``Evolving laws and cosmological energy,''
[arXiv:2306.08390 [hep-th]], to be published in Phys. Rev. D.

\bibitem{vikman}
P.~Jirou\v{s}ek, K.~Shimada, A.~Vikman and M.~Yamaguchi,
%``Losing the trace to find dynamical Newton or Planck constants,''
JCAP \textbf{04}, 028 (2021).
%doi:10.1088/1475-7516/2021/04/028
%[arXiv:2011.07055 [gr-qc]].

\bibitem{vikman1}
A.~Vikman,
%``Global Dynamics for Newton and Planck,''
[arXiv:2107.09601 [gr-qc]].

%\cite{Hammer:2020dqp}
\bibitem{vikman2}
K.~Hammer, P.~Jirousek and A.~Vikman,
%``Axionic cosmological constant,''
[arXiv:2001.03169 [gr-qc]].
%20 citations counted in INSPIRE as of 07 Nov 2023

\bibitem{Hawking3f}
S.~W.~Hawking,
%``The Cosmological Constant Is Probably Zero,''
Phys. Lett. B \textbf{134}, 403 (1984)
doi:10.1016/0370-2693(84)91370-4

\bibitem{JoaoLetter}
J.~Magueijo,
%``Cosmological time and the constants of nature,''
Phys. Lett. B \textbf{820}, 136487 (2021)
doi:10.1016/j.physletb.2021.136487
[arXiv:2104.11529 [gr-qc]].

\bibitem{JoaoPaper}
J.~Magueijo,
%``Connection between cosmological time and the constants of nature,''
Phys. Rev. D \textbf{106}, no.8, 084021 (2022)
doi:10.1103/PhysRevD.106.084021
[arXiv:2110.05920 [gr-qc]].


\bibitem{gielen} 
S.~Gielen and L.~Men\'endez-Pidal,
%``Singularity resolution depends on the clock,''
Class. Quant. Grav. \textbf{37}, no.20, 205018 (2020).
%doi:10.1088/1361-6382/abb14f
%[arXiv:2005.05357 [gr-qc]].

\bibitem{gielen1}
S.~Gielen and L.~Men\'endez-Pidal,
%``Unitarity, clock dependence and quantum recollapse in quantum cosmology,''
Class. Quant. Grav. \textbf{39}, no.7, 075011 (2022)
doi:10.1088/1361-6382/ac504f
[arXiv:2109.02660 [gr-qc]].

%\cite{Singleton:2016yal}
\bibitem{mach_conflict}
D.~Singleton and S.~Wilburn,
%``Global versus local \textemdash{} Mach\textquoteright{}s principle versus the equivalence principle,''
Int. J. Mod. Phys. D \textbf{25}, no.12, 1644009 (2016)
doi:10.1142/S0218271816440090
[arXiv:1607.01442 [gr-qc]].
%9 citations counted in INSPIRE as of 13 Nov 2023




%\cite{SupernovaSearchTeam:1998fmf}
\bibitem{Supernova}
A.~G.~Riess \textit{et al.} [Supernova Search Team],
%``Observational evidence from supernovae for an accelerating universe and a cosmological constant,''
Astron. J. \textbf{116}, 1009-1038 (1998)
doi:10.1086/300499
[arXiv:astro-ph/9805201 [astro-ph]].
%15737 citations counted in INSPIRE as of 13 Nov 2023

%\cite{Papagiannopoulos:2016dqw}
\bibitem{BD_potential}
G.~Papagiannopoulos, J.~D.~Barrow, S.~Basilakos, A.~Giacomini and A.~Paliathanasis,
%``Dynamical symmetries in Brans-Dicke cosmology,''
Phys. Rev. D \textbf{95}, no.2, 024021 (2017)
doi:10.1103/PhysRevD.95.024021
[arXiv:1611.00667 [gr-qc]].
%35 citations counted in INSPIRE as of 13 Nov 2023

%\cite{Bertotti:2003rm}
\bibitem{cassini}
B.~Bertotti, L.~Iess and P.~Tortora,
%``A test of general relativity using radio links with the Cassini spacecraft,''
Nature \textbf{425}, 374-376 (2003)
doi:10.1038/nature01997
%1484 citations counted in INSPIRE as of 13 Nov 2023

%\cite{Ellis:2003pw}
\bibitem{EllisVSL}
G.~F.~R.~Ellis and J.~P.~Uzan,
%```c' is the speed of light, isn't it?,''
Am. J. Phys. \textbf{73}, 240-247 (2005)
doi:10.1119/1.1819929
[arXiv:gr-qc/0305099 [gr-qc]].
%126 citations counted in INSPIRE as of 11 Dec 2023

\bibitem{cphi}
P.~M.Bassani, in preparation.

%\cite{Brown:1992kc}
\bibitem{brown}
J.~D.~Brown,
%``Action functionals for relativistic perfect fluids,''
Class. Quant. Grav. \textbf{10}, 1579-1606 (1993)
doi:10.1088/0264-9381/10/8/017
[arXiv:gr-qc/9304026 [gr-qc]].
%337 citations counted in INSPIRE as of 11 Dec 2023

\bibitem{twot}
B.~Alexandre and J.~Magueijo,
%``Semiclassical limit problems with concurrent use of several clocks in quantum cosmology,''
Phys. Rev. D \textbf{104}, no.12, 124069 (2021)
doi:10.1103/PhysRevD.104.124069
[arXiv:2110.10835 [gr-qc]].



%\bibitem[Hartle and Hawking(1983)]{HH}J.~B.~Hartle and S.~W.~Hawking,``Wave Function of the Universe,'' {\em Phys.\ Rev.\ D} \textbf{28} (1983), 2960--2975.




\bibitem{Paolothesis}
P.~M.~Bassani,
``Varying Constants and the Brans-Dicke theory: a new landscape in cosmological energy conservation,'',
[arXiv:2312.04260 [gr-qc]].
 

%\cite{Albrecht:1998ir}
\bibitem{AM}
A.~Albrecht and J.~Magueijo,
%``A Time varying speed of light as a solution to cosmological puzzles,''
Phys. Rev. D \textbf{59}, 043516 (1999)
doi:10.1103/PhysRevD.59.043516
[arXiv:astro-ph/9811018 [astro-ph]].
%427 citations counted in INSPIRE as of 08 Nov 2022

\bibitem{MoffatVSL}
J.~W.~Moffat,
%``Superluminary universe: A Possible solution to the initial value problem in cosmology,''
Int. J. Mod. Phys. D \textbf{2}, 351-366 (1993)
doi:10.1142/S0218271893000246


\bibitem{VSLreview}
J.~Magueijo,
%``New varying speed of light theories,''
Rept. Prog. Phys. \textbf{66}, 2025 (2003)
doi:10.1088/0034-4885/66/11/R04
[arXiv:astro-ph/0305457 [astro-ph]].



%%%%%%% copied over from elsewhere, cut from here on at the end






\end{thebibliography}

\end{document}